\documentclass[twocolumn,pra,showpacs,amsmath,amssymb,amsfonts]{revtex4}

\usepackage{graphicx}
\usepackage{bbold}
\usepackage{times}
\usepackage{float}
\usepackage[h]{esvect}
\usepackage{appendix}
\usepackage{chemarr}
\usepackage{multirow}

\usepackage{subfig}

\makeatletter

\newcommand\tabcaption{\def\@captype{table}\caption}
\newcommand{\ket}[1]{\left|#1\right\rangle}

\begin{document}

\title{The effect of optical lattice heating on the momentum distribution of a 1D Bose gas}

\author{Jean-F\'{e}lix Riou, Laura A. Zundel, Aaron Reinhard\footnote{Present address: Department of Physics, Otterbein University, 1 South Grove St., Westerville, OH 43081, USA}, and David S. Weiss}
\affiliation{ Physics Department,  The Pennsylvania State University, 104 Davey Lab, University Park, PA 16802}

\date{\today}

\begin{abstract}
We theoretically study how excitations due to spontaneous emission and trap fluctuations combine with elastic collisions to change the momentum distribution of a trapped non-degenerate one-dimensional (1D) Bose gas. Using calculated collisional relaxation rates, we first present a semi-analytical model for the momentum distribution evolution to get insight into the main processes responsible for the system dynamics. We then present a Monte-Carlo simulation that includes features that cannot be handled analytically, and compare its results to experimental data. These calculations provide a baseline for how integrable 1D Bose gases evolve due to heating processes in the absence of diffractive collisions that might thermalize the gases.
\end{abstract}


\maketitle

\tableofcontents

\section*{Introduction}

An isolated integrable many-body system does not thermalize in a conventional way. Because of its large number of conserved quantities, it approaches an asymptotic state that depends on the details of its initial state, rather than approaching an equilibrium state that depends only on its energy. For a quantum many-body system, an equivalent statement is that the eigenstate thermalization hypothesis (ETH) is not in general satisfied for an integrable system. It is an interesting and open question as to how far from integrable a many-body system must be before the ETH is satisfied. Alongside considerable theoretical work \cite{Theorywork1,Theorywork2,Theorywork3,Theorywork4}, there are ongoing experimental efforts using trapped, cold atoms to shed light on this problem \cite{ExperimentalWork1,ExperimentalWork2,ExperimentalWork3}. A central experimental problem is to distinguish a system's evolution due to internal thermalization, or more precisely, to the dephasing of its off-diagonal density matrix elements, from evolution due to fluctuations associated with the trapping potential. In this paper, we present theoretical calculations that allow us to determine how gases of atoms trapped in a 2D optical lattice evolve due to the gamut of lattice-related fluctuations present in an actual experiment, assuming integrable dynamics. The ultimate goal is to be able to isolate the incipient effect of weak non-integrability in experiments.

Although our theoretical approach is fairly general, we will focus mostly on the experimental geometry of Ref. \cite{ExperimentalWork1}, the quantum Newton's cradle geometry.  A blue-detuned 2D optical lattice confines atoms in isolated tubes. The lattice is deep enough that tunneling among the tubes is negligible. Axial confinement is provided by a Gaussian-shaped trap created by an independent red-detuned beam. The axial trap depth is made sufficiently small that there is never enough energy for pairwise collisions to populate higher vibrational states of the lattice. Therefore, in the absence of heating, the dynamics are purely 1D. These gases are well-described by the Lieb-Liniger model of 1D bosons with $\delta$-function interactions \cite{deltainteractions1,deltainteractions2}. We assume that the kinetic energy of the gas far exceeds the interaction energy, so that when an atom is lost from the gas its loss does not significantly affect the momentum distribution of the atoms that remain.

Trap fluctuations and spontaneous emission are colloquially described as causing ``heating'', as we have already done above, but that term is rather fraught. It suggests the featureless deposition of energy that characterizes heating in non-integrable systems. For an integrable system, how energy is deposited strongly affects how that deposition affects observables. A more apt term than ``heating'' would be ``stochastic quantum quenches'', since trap fluctuations and spontaneous emission project the system into new states with their own, modified sets of conserved quantities. For instance, an integrable system that initially looks like it has a thermal distribution can be projected onto obviously non-thermal distributions by the action of stochastic quantum quenches. It is also possible for an initial distribution that does not look thermal to be projected onto one that does looks thermal, even when there are no thermalizing collisions. In the interest of readability, we will continue to use the term ``heating'', with the understanding that it encompasses this specialized meaning in integrable systems.

The rates associated with the effects of spontaneous emission of lattice photons have been calculated elsewhere \cite{TransitionRates,TransitionRates1,TransitionRates2}. Spontaneous emission can directly deposit axial energy, and it can transversely excite the atoms. The subsequent collisional de-excitation of transversely excited atoms is often the dominant way that spontaneous emission deposits energy into the 1D gas; the exposition of this process is the central part of this paper. Heating from fluctuations in lattice depth and position depends on experimental imperfections that are not always easy to determine directly, but they can often be empirically inferred from measurements of 1D gases. Using simplifying assumptions, we develop analytical expressions for how a 1D Bose gas' momentum distribution evolves due to each of these processes, and for the associated atom loss. We then describe a Monte-Carlo simulation that need not have those simplifying assumptions. Although the analytical expressions yield qualitative understanding, the Monte-Carlo simulations give results that differ in appreciable ways and accurately reproduce experimental data.

\vspace{1cm}

\section{Elastic collisional dynamics in a 1D tube}

Throughout this paper we consider an array of tubes aligned along the $z$-axis, each confining an independent 1D Bose gas. The external state of an atom in a waveguide is written as $\ket{\phi}\equiv\ket{n_x,n_y,p}$, where $n_x$ and $n_y$ are its transverse vibrational quantum numbers and $p$ is its axial momentum. We consider atoms that are initially in the transverse vibrational ground state $\ket{0, 0, p}$. Atoms can be heated to higher transverse vibrational levels, $\ket{n_x,n_y}$, by spontaneous emission or by mechanical lattice fluctuations. In the former case, transverse excitation  is correlated with axial photon recoils \cite{TransitionRates}, and that axial energy can be collisionally transferred to $\ket{0,0}$ atoms via two-body collisions that do not change transverse occupations. In the latter case, there need not be direct axial heating.  But in either case, the dominant way that transverse excitation energy gets coupled into axial motion is through collisions in which atoms decay back to $\ket{0,0}$.
The rates of the excitation processes associated with photon scattering and technical noise have been derived elsewhere \cite{TransitionRates,Gehm98}. In this section we will describe the collisional processes that convert transverse excitations into changes in the axial momentum distribution. We will ignore states with more than two quanta of excitations, since, as justified below, these are rarely populated and are quickly lost.

The theory of collisions in a waveguide \cite{Olshanii1998,HouchesLecture,MikeMoorePrivateCommunication} gives the probability $P_{\alpha \beta}^{\gamma \delta}(k)$ of the output of a collision
\begin{equation}
\alpha+\beta \rightarrow \gamma +\delta
\end{equation}
where $\alpha$ ($\gamma$) and $\beta$ ($\delta$)  are the quantum numbers associated with the transverse states, $\ket{n_x,n_y}$, of the two particles before (after) the collision, with initial momenta $p_1$ and $p_2$. These probabilities depend upon the waveguide confinement, and are parameterized by $\left(k a_{\perp}\right)^2$ where $k$ is the wavevector of the relative motion $\hbar k=(p_1-p_2)/2$, with mass $\mu=m/2$, and where $a_{\perp}=\sqrt{\hbar/\mu \omega_{\mathrm{latt}}}$ is the transverse harmonic oscillator length in the harmonic approximation.

\subsection{Absence of collisional excitation from $\ket{n_x=0,n_y=0}$}

Not all transverse states are connected via two-body collisions. For example, the following collisional processes
\begin{equation}
\ket{0,0}+\ket{0,0} \rightarrow
\left\{
\begin{gathered}
\ket{1,0}+\ket{0,0}\\
\ket{0,1}+\ket{0,0}
\end{gathered}
\right.
\end{equation}
are forbidden by parity selection rules that prevent the number of transverse quanta from changing by an odd number. The processes
\begin{equation}
\ket{0,0}+\ket{0,0} \rightarrow
\left\{
\begin{gathered}
\ket{2,0}+\ket{0,0}\\
\ket{1,0}+\ket{1,0}\\
\ket{1,0}+\ket{0,1}\\
\ket{0,2}+\ket{0,0}
\end{gathered}
\right.
\end{equation}
are parity allowed, but forbidden by energy conservation given the central premise of the experimental design, that there are no pairwise collisional excitations from $\ket{0,0}$ to higher states. In the harmonic limit, this 1D condition can be stated as $U_0\leq \hbar \omega_{\mathrm{latt}}$, where $U_0$ is the axial trap depth, or $\left(k a_{\perp}\right)^2\leq4$,
which defines the range of $k a_{\perp}$ available in the $\ket{0,0}$ state.

\subsection{Relevant collisional processes for the quantum Newton's cradle}

If we consider the experimental configuration of \cite{ExperimentalWork1}, we notice that the following state-changing collisional processes are parity and energetically allowed:
\begin{eqnarray}
\ket{1,0}+\ket{0,1}&\xrightleftharpoons[]{}&\ket{1,1}+\ket{0,0} \label{procexchange1}\\
\left.
\begin{gathered}
\ket{1,0}+\ket{1,0}  \\
\upharpoonleft\!\downharpoonright \\
\ket{0,1}+\ket{0,1}
\end{gathered}
\right\}&\xrightleftharpoons[]{}&
\left\{
\begin{gathered}
\ket{2,0}+\ket{0,0}   \\
\upharpoonleft\!\downharpoonright\\
\ket{0,2}+\ket{0,0}
\label{1010-2000}
\end{gathered}
\right. \label{procexchange2} \\
\searrow&&\swarrow \notag\\
&\ket{0,0}+\ket{0,0}& \label{procdecay}
\end{eqnarray}

The processes enumerated in Eqs. \ref{procexchange1} and \ref{procexchange2} are reversible, occurring with a probability $P_{\alpha \beta}^{\gamma \delta}(k)$. The transition rates $\Gamma_{\alpha \beta}^{\gamma \delta}$ depend upon the state populations (see eq. \ref{eqGammaNiNj}). The matrix elements involved in the de-excitation processes of Eq. \ref{procdecay} are also, of course, inherently reversible, but atoms are highly likely to be lost out of the ends of the 1D tubes before the corresponding excitation occurs, which is illustrated by the unidirectional arrows for those cases.

\begin{figure}[H]
\begin{center}
\includegraphics[width=3.4in, angle=0]{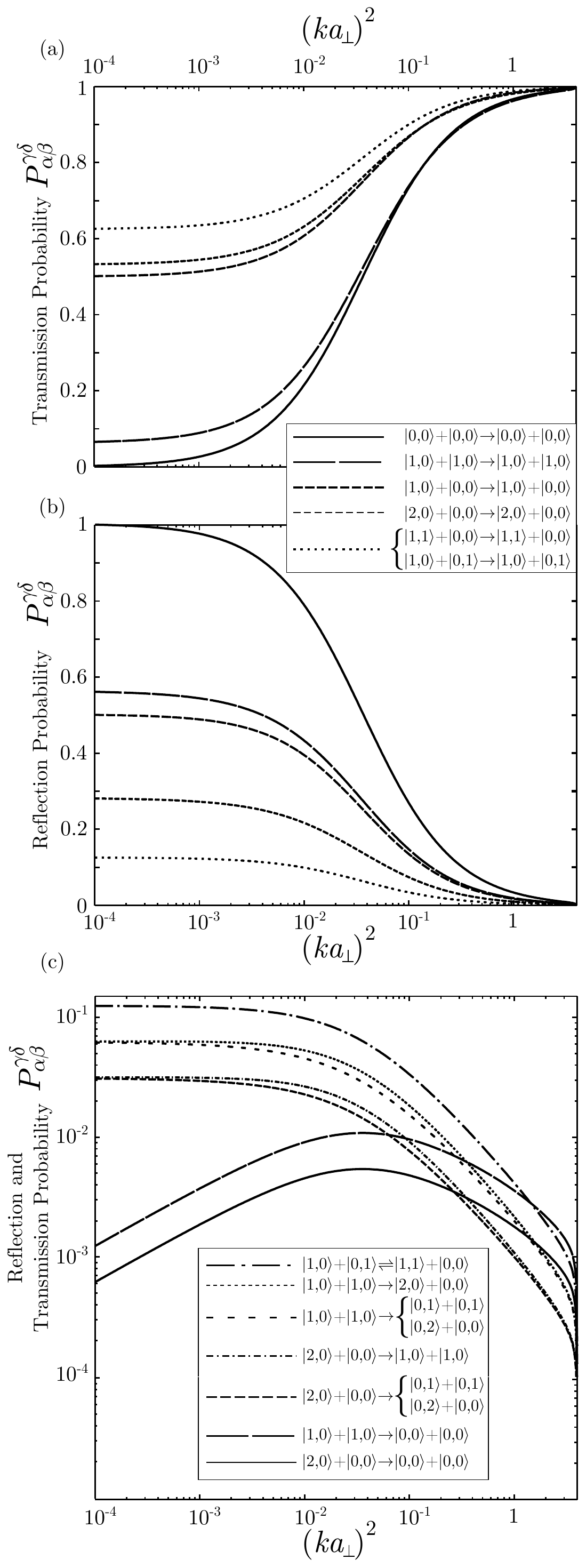}
\caption{Probabilities $P_{\alpha \beta}^{\gamma \delta}$ associated with the processes of interest for $(k a_{\perp})^2 \leq 4$ and a lattice depth of 60 $E_r$. For the six processes listed in the top inset, we plot (a) the transmission probabilities and (b) the reflection probabilities. (c) Case of nine processes for which the transmission and reflection probabilities are the same.}
\label{PABGD}
\end{center}
\end{figure}

The atomic population remains mainly in the ground state, so we treat the population in the excited state perturbatively. The population in the $\mathrm{n^{th}}$ excited state is of the $\mathrm{n^{th}}$ order in perturbation; we neglect $n$ above two.  We also restrict our analysis to two-body collisions, since higher order collision probabilities are down by at least one perturbative order. If the interactions are integrable, the higher order collisions will be relatively unimportant. If the interactions are not fully integrable, then higher order collisions might be diffractive and could thermalize the system, in contrast to the collisions considered here.

Figure \ref{PABGD} shows  $P_{\alpha \beta}^{\gamma \delta}$ for the collisions considered here. A normalization condition holds, $\sum_{\gamma \delta} P_{\alpha \beta}^{\gamma \delta}=1$. The plot is made for ${}^{87}\mathrm{Rb}$ in a lattice depth of 60 $E_r$ ($a_\perp=62.7$ nm). These results generalize the transmission probabilities for atoms in the ground state of a waveguide \cite{Olshanii1998}. The de-excitation probabilities that are most important in this work, like that for $\ket{1,0}+\ket{0,1} \rightarrow \ket{0,0}+\ket{0,0}$, are smaller than ground state reflection probabilities over the entire range of available $k a_{\perp}$.

\section{General framework for calculating the effect of heating on the momentum distribution}
\label{framework}

The collisions enumerated in the previous section all require at least one incoming atom to be in a transversely excited state. In the 1D experimental geometry they can get there only by spontaneous emission \cite{TransitionRates} or technical noise \cite{Gehm98}. These processes can also deposit energy directly into axial motion. In this section, we will develop a general framework for calculating how various heating processes affect the axial momentum distribution $f(p)$ and the loss of atoms, and how much net energy they end up depositing in 1D gases.

\subsection{$p_0$ and $E_0$ representation of the momentum distribution}

In a quantum Newton's cradle experiment, atoms are diffracted in 1D so that they start oscillating in a nearly harmonic axial trap in two 180 degrees out of phase groups. Each group has a finite spatial extent, and also undergoes a breathing motion, since their internal repulsion is not balanced by trap curvature. There is thus an inevitable spread in peak amplitudes across the atom distribution, which evolves as the interactions among atoms changes. After several cycles, the number of which depends on the degree of anharmonicity (it is 8 to 15 in our experiments), the phases of the individual oscillating atoms are no longer correlated. That is, the oscillations are dephased. The oscillation phase is the unconstrained dynamical variable, which is called the angle variable in classical mechanics. What we call ``dephasing'' in this case has been called ``prethermalization'' in other contexts \cite{Berges,ExperimentalWork3}. When it is complete, dynamical evolution ceases, unless there are thermalizing collisions or quantum quenches (like heating).

Once the quantum Newton's cradle is dephased, the conserved quantity (which corresponds to the action variable in classical mechanics) is the mechanical energy, $E_0=p_0^2/2m$, where $p_0$ is the amplitude in momentum space. It is natural to decompose the momentum distribution $f(p)$, which is what is typically measured, onto the amplitude basis $p_0$, i.e., onto the basis of dephased functions of constant mechanical energy $E_0$,
\begin{equation}
{}^{}\!\!\!\!\!f(p)\!=\!\!\int_0^{\sqrt{2mU_0}}\!\!\!\!\!\!\!\!dp_0\,G(p,p_0)f(p_0)\!=\!\!\int_0^{U_0}\!\!\!\!\!dE_0\,G(p,E_0)f(E_0).
\label{eq:fp0-p}
\end{equation}
$G(p,p_0)$ is the time-averaged momentum distribution of atoms with $p_0$, where $0 \leq p \leq p_0$. It is also the instantaneous momentum distribution of an ensemble of dephased atoms with peak momentum $p_0$, which is what is relevant for the steady state in a quantum Newton's cradle. Its precise functional form depends on the axial potential $U(z)$ of depth $U_0$. The elements of the two integrals in Eq. \ref{eq:fp0-p} can be readily related to each other, since $f(E_0)dE_0=f(p_0)dp_0$ leads to $f(E_0)=m f(p_0)/p_0$ and $G(p,E_0)\equiv G(p,p_0=\sqrt{2mE_0})$.
The momentum distribution $f(p)$ directly gives the kinetic energy $E_{\mathrm{kin}}=\int dp \,\,\frac{p^2}{2m} f(p)$, whereas $f(p_0)$ or $f(E_0)$ give access to the total mechanical energy $E_{\mathrm{mec}}=\int dp_0 \,\,\frac{p_0^2}{2m} f(p_0)=\int dE_0 \,\,E_0 f(E_0)$. Note that $f(p_0)$ can be obtained from $f(p)$ by numerically inverting the integral in Eq. \ref{eq:fp0-p}.

\subsection{The heating point-spread function in energy space}

Each heating mechanism we consider (labeled by $(i)$) is fully characterized in our model by 2 quantities:
a rate $R^{(i)}$ at which it happens; and
a function $F^{(i)}(p_0,p,\Delta p)$, which gives the probability that one event transfers a momentum kick $\Delta p$ to an atom with mechanical energy $E_0$ and momentum $p$. This function is normalized, $\int d\Delta p \,F^{(i)}(p_0,p,\Delta p)=1$.

Heating will cause the momentum distribution to evolve in time, $f(p,t)$. In this section we will describe how to obtain $f_{\mathrm{h}}(p,t+\Delta t)$, the heated distribution after an elementary time step $\Delta t$. For one specific process $(i)$, each heating event gives an instantaneous momentum kick $\Delta p$ to a single atom. The corresponding change $\Delta E_0$ of its mechanical energy is from the change of its kinetic energy since the atom's position along the tube does not change during the event, i.e. $\Delta E_0 =(2 p \Delta p +\Delta p^2)/2m$.

We need to know, given $E_0$, the probability $P^{(i)}(E_0,\Delta E_0)\, d\Delta E_0 $ that an atom experiences a given $\Delta E_0$ after one heating event. We build this distribution by first considering the joint probability density $P^{(i)}_j(p_0,p,\Delta p)$ that, at a given time, an atom has a momentum $p$ and gets a momentum kick $\Delta p$. Since $p$ and $\Delta p$ are independent variables,
\begin{equation}
P^{(i)}_j(p_0,p,\Delta p)= G(p,p_0) F^{(i)}(p_0,p,\Delta p),
\end{equation}
where $G(p,p_0)dp$ is the probability that an atom with $p_0$ has momentum $p$, and $F^{(i)}(p_0,p,\Delta p)d\Delta p$ is the probability that an atom with $p_0$ and $p$ experiences a given $\Delta p$.

Performing the change of variable $p \rightarrow \Delta E_0$ while keeping $\Delta p$ constant and integrating over all momentum changes $\Delta p$, we find the probability that an atom with energy $E_0$ has its energy changed by $\Delta E_0$ to be
\begin{equation}
P^{(i)}(E_0,\Delta E_0)\!\!=\!\!\int \frac{m}{|\Delta p|} G(p,p_0) F^{(i)}(p_0,p,\Delta p)\,d\Delta p
\end{equation}
where $p_0\!=\!\sqrt{2mE_0}$ and $p=(2m\Delta E_0 -\Delta p^2)/2 \Delta p$. The point spread function $P^{(i)}$ is normalized in energy space, $\int_{-E_0}^{\infty}d\Delta E_0 \,\,P^{(i)}(E_0,\Delta E_0)=1$.

\subsection{Distribution evolution}

\begin{figure}
\begin{center}
\includegraphics[width=3.4in, angle=0]{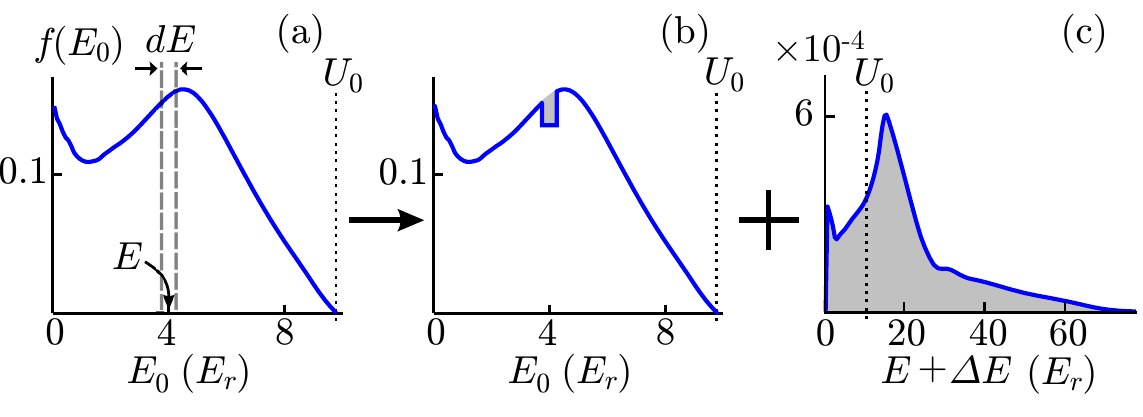}
\caption{Impact of heating on a class of atoms with a mechanical energy $E$ within $dE$ in an initial distribution $f(E_0)$ (fig. a). The heating applied on this class results in the distribution sum of (b) and (c).  (b) The initial distribution is depleted at $E$ with a  proportion $R^{(i)} \Delta t$ of atoms (grey area). (c) The energy of these atoms is spread according to the point spread function $P^{(i)}(E,\Delta E)$. (The total grey area under the curve is the same as that in figure b). Only the atoms with a final energy $E+\Delta E$ smaller than the axial depth $U_0$ contribute to the net heating, and the other atoms are lost.  We have plotted these curves for an energy $E=4\,E_r$. The process illustrated here is the heating when collisional de-excitation releases energy $E_g=22.5\,E_r$, which is the case of a lattice depth of 40 $E_r$.}
\label{labelExplainFramework}
\end{center}
\end{figure}

Each specific heating process $(i)$ has an associated point spread function $P^{(i)}(E_0,\Delta E_0)$.  We obtain the heated distribution $f^{(i)}_H(p)$ in $p$ space by applying the transformation $E_0\rightarrow p$ via $G(p,E_0)$ and the point spread function (see fig. \ref{labelExplainFramework}):
\begin{equation}
{}^{}\!\!\!\!f^{(i)}_H(p)\!=\!\!\!\int_0^{U_0}\!\!\!\!\!\!\!dE_0\,G(p,E_0)\!\!\!\int_{-E_0}^{U_0-E_0}\!\!\!\!\!\!\!\!\!\!\!\!\!\!\!d\Delta E_0\,f(E_0)P^{(i)}(E_0,\Delta E_0)
\label{eq:fiHp}
\end{equation}
 The lower limit in the $\Delta E_0$ integral is set by the atom's mechanical energy. Note that $\int dp f^{(i)}_H(p)\leq 1$, since heating can promote an atom to an energy which exceeds the trap depth (when $E_0+\Delta E_0 \geq U_0$), in which case that atom (or an identical atom with which it exchanges energy and momentum) will be lost within half an axial oscillation period.

During a time step $\Delta t$, a proportion of atoms $R^{(i)} \Delta t$ participates to the heating. Over that time step, $1-R^{(i)}\Delta t$ of the distribution is unchanged. The distribution at the end of the time step is the sum of the unchanged part of the distribution and the heating modified part of the distribution,
\begin{equation}
f_{\mathrm{h}}(p,t\!+\!\Delta t)\!=\!\left[\!1\!-\!\Delta t\!\sum_i\!R^{(i)}\!\right]\!f(p,t)\!+\!\Delta t\!\sum_i\!R^{(i)}\!f^{(i)}_{H}(p,t)
\label{geneframework}
\end{equation}

\subsection{Associated losses}
The fraction of atoms that are lost because their mechanical energy exceeds the trap depth is given by
\begin{equation}
\frac{N(t+\Delta t)}{N(t)}=1-\Delta t \sum_i \widetilde{c}^{(i)} R^{(i)} .
\label{eqlossfH}
\end{equation}
where $\widetilde{c}^{(i)}$ is the ratio of the excitation rate to the associated loss rate,
\begin{equation}
\widetilde{c}^{(i)}=1-\int dp f^{(i)}_H(p)
\label{eqctildei}
\end{equation}
This coefficient can be calculated by applying the heating process to a particular momentum distribution. In general, the loss rate depends on the momentum distribution.

\subsection{Deposited energy}
$E_{\mathrm{dep}}^{(i)}$ is the average energy deposited axially for the process $(i)$. It is given by
\begin{equation}
E_{\mathrm{dep}}^{(i)}=\int dp_0 \frac{p_0^2}{2m} \left[f^{(i)}_{H}(p_0) - \left(1-\widetilde{c}^{(i)}\right)f(p_0)\right]
\end{equation}
where $f(p_0)$ and $f^{(i)}_{H}(p_0)$ are the $p_0$ representations of $f(p)$ and $f^{(i)}_H(p)$.

\section{Application to a 1D gas trapped in a transverse 2D blue-detuned optical lattice and in an axial gaussian beam}

 In this section, we apply the previous formalism to the quantum Newton's cradle experiment as described in Ref. \cite{ExperimentalWork1}.

\subsection{Trap configuration}
We consider an atomic gas trapped in $x$ and $y$ at the node of a deep 2D optical lattice. This lattice is made using two perpendicular 1D lattices with $k$-vector, $k_r$, each with a slightly different frequency so that the two 1D lattices do not mutually interfere \cite{Winoto}. Each 1D lattice is linearly polarized perpendicular to the axial direction $z$. For atoms that do not change hyperfine state, the spontaneous emission rate at the nodes of the 2D blue lattice is $2\Gamma^{\mathrm{blue}}$. Axial trapping is provided by an additional trap of gaussian profile with depth $U_0$ and waist $w_0$, $U(z)=U_0\left[1-\exp\left(-2 z^2/w_0^2\right)\right]$. The axial trapping is weak enough that motion in that direction can be treated semi-classically and characterized by the continuous momentum variable $p$. The 2D lattice is deep enough that tunneling among tubes can be neglected. Atoms start initially in the transverse ground state $\ket{n_x,n_y}=\ket{0,0}$. Some of them are temporarily transferred to higher transverse states as a result of spontaneous emission or lattice fluctuations.

\subsection{The dephased momentum distribution}
\label{refp0tilte}

\begin{figure}
\begin{center}
\includegraphics[width=3.4in, angle=0]{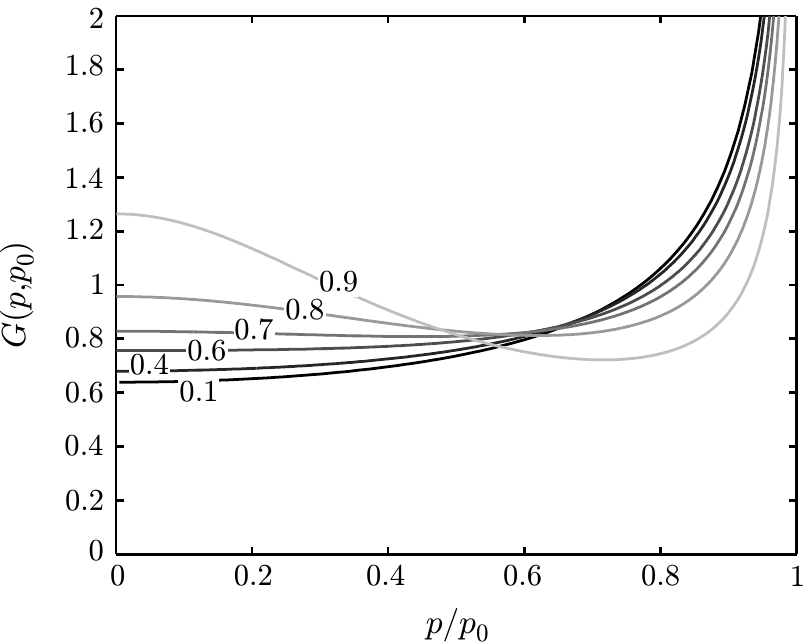}
\caption{Plot of $G(p,p_0)$ (Eq. \ref{eq:Gpp0}). The family of curves is parameterized by $\tilde{p}_0=\{0.1,\, 0.4,\, 0.6,\, 0.7,\, 0.8,\, 0.9\}$. As $\tilde{p}_0$ increases, the function deviates strongly from the harmonic limit (which is indistinguishable from $\tilde{p}_0=0.1$)  and the atom spends more time at the edges of the Gaussian trap. This increases the relative weight of small momenta $p$ in the distribution. }
\label{figGppo}
\end{center}
\end{figure}

For a Gaussian potential of depth $U_0$, the dephased momentum distribution $G(p,p_0)$ is given by
\begin{equation}
G(p,p_0) = \frac{\mathrm{Norm(}\widetilde{p_0})}{\sqrt{-\ln(1-\widetilde{p_0}^2+\widetilde{p_{}}^2)}(1-\widetilde{p_0}^2+\widetilde{p_{}}^2)}
\label{eq:Gpp0}
\end{equation}
where we have used the reduced momenta $\widetilde{p_0}=p_0/\sqrt{2 m U_0}$ and  $\widetilde{p}=p/\sqrt{2 m U_0}$. This general expression is derived in Appendix \ref{AppGpp0}, along with the particular result for the case of harmonic trapping. The normalization prefactor $\mathrm{Norm}(\widetilde{p_0})$ is calculated numerically and tabulated. This function is plotted for different values of $\tilde{p}_0$ in Fig. \ref{figGppo}.

\subsection{Excitation processes and heating mechanisms in the system}
\label{consideredprocesses}

As described in Ref. \cite{TransitionRates}, spontaneous emission in the lattice can lead to axial deposition of energy with or without a change of transverse state $\ket{n_x,n_y}$.
It can also change an atom's magnetic sublevel, which can modify the momentum distribution if different sublevels are subject to different axial potentials, as is the case in \cite{ExperimentalWork1} where a levitating magnetic field gradient is used. The deposition of energy associated with this last process cannot be taken into account in our semi-analytical framework, and must be handled in a Monte-Carlo simulation. For all of these processes, the transition rates and the probability distributions $F$ for getting a given axial momentum kick $\Delta p$ are calculated in \cite{TransitionRates} from first principles using only the lattice parameters. 

Collisions such as $\ket{1,0}+\ket{0,0}\rightarrow \ket{1,0}+\ket{0,0}$, do not change the internal state but can lead to an exchange of the momenta of the two atoms. This means that there is a fast equilibration of momentum distributions of different $\ket{n_x,n_y}$ states. Spontaneous emission creates a correlation between the axial momentum and the transverse excitations. However, we neglect it here because of this fast equilibration, which is also observed in the Monte-Carlo simulation.

Excitations associated with technical noise in conservative traps are described in \cite{Gehm98}. The associated transition rates can be calculated from the power spectrum of the noise. Noise from lattice position or the lattice intensity fluctuations can drive transitions that change the $x$ or $y$ transverse state by either $n=$1 or 2 quanta at the rate $R_{0\rightarrow n}$. Noise associated with axial trap motion or axial confinement fluctuations cannot be handled in our semi-analytical model, but are included in the Monte-Carlo simulation.

According to Eq. \ref{geneframework}, the evolution of the momentum distribution $f(p)$ is completely determined by the rate $R^{(i)}$ at which each heating mechanism happens and by the associated probability distribution $F(p_0,p,\Delta p)$ for getting a given momentum kick $\Delta p$ along $z$.
We can split all the heating mechanisms in two categories: those that do involve a collision and those that do not. For the conditions of Ref. \cite{ExperimentalWork1}, we will present the leading order terms in $R^{(i)}$ and $F^{(i)}$.

\subsection{Mechanisms involving a de-excitation collision}
\label{sec:deeccol}

When an atom is excited into a higher transverse state, most of the effect on the axial momentum distribution only occurs after binary collisions like those described in Eq. 6 lead to transverse to axial energy conversion and atom loss. If we assume that the system is in quasi steady-state, then the collisional de-excitation rate $R^{(i)}$ per atom in $\ket{0,0}$ equals the sum of the excitation rates $\Gamma_{00}^{n'_xx'_y}$ from the ground state $\ket{0,0}$ to higher excited states $\ket{n'_xx'_y}$, where $n'_x+n'_y \leq 2$. That is, $R^{(i)}= \Gamma_{00}^{10}+\Gamma_{00}^{01}+2\Gamma_{00}^{11}+2\left(\Gamma_{00}^{20}+\Gamma_{00}^{02}\right)$. The first factor of two comes from the fact that an excitation of one atom in the two orthogonal transverse directions is indistinguishable from two atoms being excited in orthogonal transverse directions.  In neither case will those excitations interact with each other. The last factor of two comes from the fact that, in the harmonic approximation, an atom in $\ket{2,0}$ can collide with an atom in $\ket{0,0}$ to give two atoms in $\ket{1,0}$. Therefore we can treat a two quanta excitation of one atom as one quantum excitation in each of two atoms. To leading order in $E_r/\hbar \omega_{\mathrm{latt}}$,
\begin{equation}
R^{(i)}=2\Gamma^{\mathrm{blue}}+2R_{0\rightarrow1}+4R_{0\rightarrow2}
\label{eqRicolllatt}
\end{equation}

To derive the function $F^{(i)}(p_0,p,\Delta p)$ for this mechanism, consider two atoms with relative momentum $2 k$, whose transverse quantum numbers are changed by a collision. By conservation of momentum and energy, the momentum kick $\Delta p$ to the atoms must satisfy
\begin{equation}
E_g=\frac{2 k \Delta p + \Delta p^2}{m}
\label{eqEgk}
\end{equation}
Given $E_g$ and $k$, the distribution associated with the $k$-dependent probability that an atom with center of mass momentum $\pm k$ gets a  momentum kick $\Delta p$, is the sum of four terms
\begin{equation}
{}^{}\!\!\!\!F_1(\Delta p,k)\!=\!\frac{1}{4}\!\!\sum_{i=1,2\atop j=1,2}\!\!\delta\!\!\left[\!\Delta p\!+\!(-1)^i k\!+\!(-1)^j \sqrt{k^2\!+\!m E_g}\right]\label{kdepF}
\end{equation}
where $\delta$ is the Dirac function.

The function $F$ is the integral of the $k$-dependent probability in Eq. \ref{kdepF} weighted by the probability function $g_{\mathrm{rel}}$ that an atom with a given $p_0$ and $p$ experiences a collision involving a relative momentum $\hbar k$ (Eq. \ref{grel}), $F(p_0,p,\Delta p)=\int\,dk\, F_1(\Delta p,k) g_{\mathrm{rel}}(p_0,p,k)$. After performing the integral, we obtain
\begin{equation}
F(p_0,p,\Delta p)=\frac{1}{4} \sum_{i=1,2\atop j=1,2} \frac{g_{\mathrm{rel}}(p_0,p,(-1)^i k_s)}{\left|1+(-1)^j k_s/\sqrt{k_s^2+mE_g} \right|}
\label{Fcoll}
\end{equation}
where $k_s(E_g,\Delta p)$ is the  solution of Eq. \ref{eqEgk} for $k$.

Although spontaneous emission does not depend on $p$ and $p_0$ (see Eqs. \ref{eq:spont1} and \ref{eq:spont2}), the distribution $F$ does, which means that the ultimate effect of spontaneous emission on heating depends on $p$ and $p_0$.  Note also that $F$ does not depend on whether the transverse excited states were populated by spontaneous emission or technical lattice noise.

\subsection{Mechanisms involving no de-excitation collision}

Although energy deposition from spontaneous emission often proceeds through transverse excitation, it can also deposit energy directly in the axial direction, leaving the atom in $\ket{n_x=0,n_y=0}$. Using the results of \cite{TransitionRates}, we calculate the parameters for this process with a blue detuned 2D lattice:
\begin{equation}
{}^{}\!\!\!R^{(i)}=\frac{3 E_r}{5 \hbar \omega_{\mathrm{latt}}}\Gamma^{\mathrm{blue}};\,F^{(i)}\!=\!\frac{5}{8}\!\left[1\!-\!\left(\frac{\Delta p}{\hbar k_r}\right)^{\!4}\right]
\label{eq:spont1}
\end{equation}

Atoms that have been previously excited to a higher transverse state can return to the ground state after a de-exciting spontaneous emission event, which involves an axial momentum kick. The rate of this mechanism {\it per atom in} $\ket{0,0}$ is given by the product of the transition rate induced by spontaneous emission from $n=1$ to $n=0$ for one direction ($x$ or $y$) and the ratio between the total population $N_1$ in $\ket{1,0}$ or $\ket{0,1}$ and the ground state population $N_0$. Applying the results of \cite{TransitionRates}, we find that the parameters for this process are
\begin{equation}
R^{(i)}=\frac{N_1}{N_0}\Gamma^{\mathrm{blue}};F^{(i)}\!=\!\frac{3}{8}\left[1+\left(\frac{\Delta p}{\hbar k_r}\right)^{\!2}\right]
\label{eq:spont2}
\end{equation}

It is worth noting that for the momentum distributions that we consider in this paper (see section \ref{sec:paramexp}), the ratio $N_1/N_0$ is typically in the range of $5-15 \%$ \cite{footnoteN1}, and $E_r/\hbar \omega_{\mathrm{latt}} \sim 1/10$. Therefore heating from direct excitation in the ground state (eq. \ref{eq:spont1}) and from de-excitation from $n=1$ via spontaneous emission (eq. \ref{eq:spont2}) have the same order of magnitude. However, as shown by eq. \ref{eq:spont1} and \ref{eq:spont2}, both processes discussed in this paragraph have a contribution to the total evolution that is negligible compared to the contribution of the de-excitation collisions, which happen at a much faster rate (eq. \ref{eqRicolllatt}).

\vspace{1 cm}
Using the results of this general framework, we have calculated the evolution of a particular initial momentum distribution, while keeping track of the energy increase and atom loss due to the heating processes. These results are summarized in figure \ref{FigA}, where they are compared to the results of Monte-Carlo simulations.

\section{Monte-Carlo simulation}
\subsection{General description}

To include in the dynamics effects that were neglected in the analytical model, we have performed a Monte-Carlo simulation of the evolution of the system. This simulation takes into account all processes at the microscopic level. It is a semi-classical calculation. Each atom is described by the product of a transverse quantum state $\ket{n_x,n_y}$ and a classical axial momentum $p$ and position $z$. The atoms interact via two-body collisions according to quantum mechanical probabilities. In the absence of collisions, the time evolution of the axial momentum and position is given by the classical equation of motion $dp/dt=-dU(z)/dz$. A two-body collision is recorded every time two atomic trajectories cross each other. The outcome of this collision can change the momenta and/or the transverse states of the two atoms with quantum probabilities given by the results of Refs. \cite{HouchesLecture} and \cite{MikeMoorePrivateCommunication}.

Each atom is also subject to transitions between different states $\ket{n_x,n_y,p}\rightarrow\ket{n'_x,n'_y,p'}$ induced by spontaneous emission due to the lattice. These processes happen randomly with rates and quantum mechanical probabilities that are calculated in \cite{TransitionRates}. In particular, these include the possibility that an atom changes hyperfine sublevel, which makes it experience a different axial potential $U(z)$ because the magnetic field gradient no longer cancels gravity.

The simulation also includes the effect of losses, independent of their source. Losses can result from background gas collisions or as a result of transitions to non-trapped states. The simulation can also include the possible impact of any technical noise on trap depth and position.

This numerical approach keeps track of the fate of each individual atom position, momentum and transverse state as a function of time, which gives access to the time evolution of all the observables of the system such as the momentum distribution, the populations in each state,  the loss rate and the energy. This is done at the single tube level as well as at the global level by averaging over different tubes to represent as faithfully as possible the experiment. 

\subsection{Outline of the simulation}

We present here the main steps in the simulation, and enumerate the assumptions of this numerical treatment.

\subsubsection{Preparation of the initial state}

We first distribute the total number among independent tubes in an effort to match the experimental distribution described in section \ref{sec:paramexp}. The maximum momentum $p_0$ of each atom is randomly chosen from the momentum distribution measured as an experimental average over all tubes. The initial positions $z$ and momenta $p$ are randomly drawn from the trajectory associated with $p_0$. We then let the system evolve under the effect of collisions and transverse (de)-excitations, but with momentum changes disabled. Atoms then redistribute themselves among the transverse states $\ket{0,0}$, $\ket{1,0}$, $\ket{0,1}$, $\ket{1,1}$, $\ket{2,0}$, $\ket{2,1}$, $\ket{2,2}$,  $\ket{1,2}$, and $\ket{0,2}$. Once the system has reached a steady state, the equilibrium transverse populations reflect the different transition and collision rates, along with number-dependent tube-to-tube variations.

\subsubsection{Evolution of the system}

At $t=0$ we let the momenta evolution proceed. Time is discretized with a variable time step determined by the time between significant collisions. A two-body collision between identical $\ket{0,0}$ atoms is insignificant, since the final states are indistinguishable. A significant collision is one in which the outcome is not trivial, i.e. which involves at least one atom in a state different from $F=1, m_F=1,\ket{0,0}$.  At every iteration, the algorithm finds the time of the next significant collision and calculates its outcome. Also, the algorithm stochastically applies transverse excitations and de-excitations and losses to the system. Atoms that experience a transition to states where $n_x$ or $n_y$ is larger than or equal to three are considered to be lost and are removed from the system, as are atoms which move away from the center of the axial trap by a distance of three times the waist of the gaussian potential $U(z)$.

\subsubsection{Assumptions}

The following features are not captured by this treatment.

\begin{itemize}
\item We do not simulate the evolution of the full quantum many body wavefunction, including n-body correlations. Such a treatment is significantly out of reach numerically, but we expect to capture the essence of the heating with this semi-classical model.
\item We do not take into account changes in the mean field energy due to losses and changes in the momentum distribution. In the quantum Newton's cradle, the vast majority of the mean field energy is released during the dephasing process, which significantly lowers the atom density.
\item We do not consider collisions involving more than 2 atoms. This means that we do not include inelastic collisions, which can in general lead to experimentally observable loss, but not for the low density distributions that we study in this paper. Higher body elastic collisions could potentially lead to thermalization and evaporation. Attempting to include them is beyond the scope of this work.
\item We assume that atoms with $n_x$ or $n_y$ larger than or equal to three are lost. For realistic experimental parameters, the population in these states is undetectable, and the tunneling between tubes in these transverse states prevents them from being trapped for long.
\item We neglect tunneling in higher bands by considering independent tubes for all $n_x$ and $n_y$. This would be problematic at lattice depths lower than thoses used in our experiment. 
\end{itemize}

\subsection{Experimental distribution to compare to the Monte-Carlo simulation}
\label{sec:paramexp}

	Our reason for studying quantum Newton's cradles \cite{ExperimentalWork1}  is to understand what their long time steady state is and how it is approached, which requires the action of 3-body diffractive collisions \cite{BeautifulModels}. The purpose of our Monte-Carlo model is to understand the net effect of  all processes other than 3-body collisions, either diffractive or inelastic \cite{Burt}. The idea will be to extract the heating from the evolution in order to isolate the 3-body effects. Although that procedure is beyond the scope of this paper, we do want to empirically test the validity of the Monte-Carlo simulation here. To do so we need to compare it to experimental distributions whose evolution is driven only by possible heating effects. Quantum Newton's cradle distributions are typically too dense to meet this criterion, and potentially have atom-atom correlations that are not included in the model. We have therefore developed the following procedure for creating a low density thermal distribution of atoms in 1D.
	
\begin{figure}[H]
\begin{center}
\includegraphics[width=3.4in, angle=0]{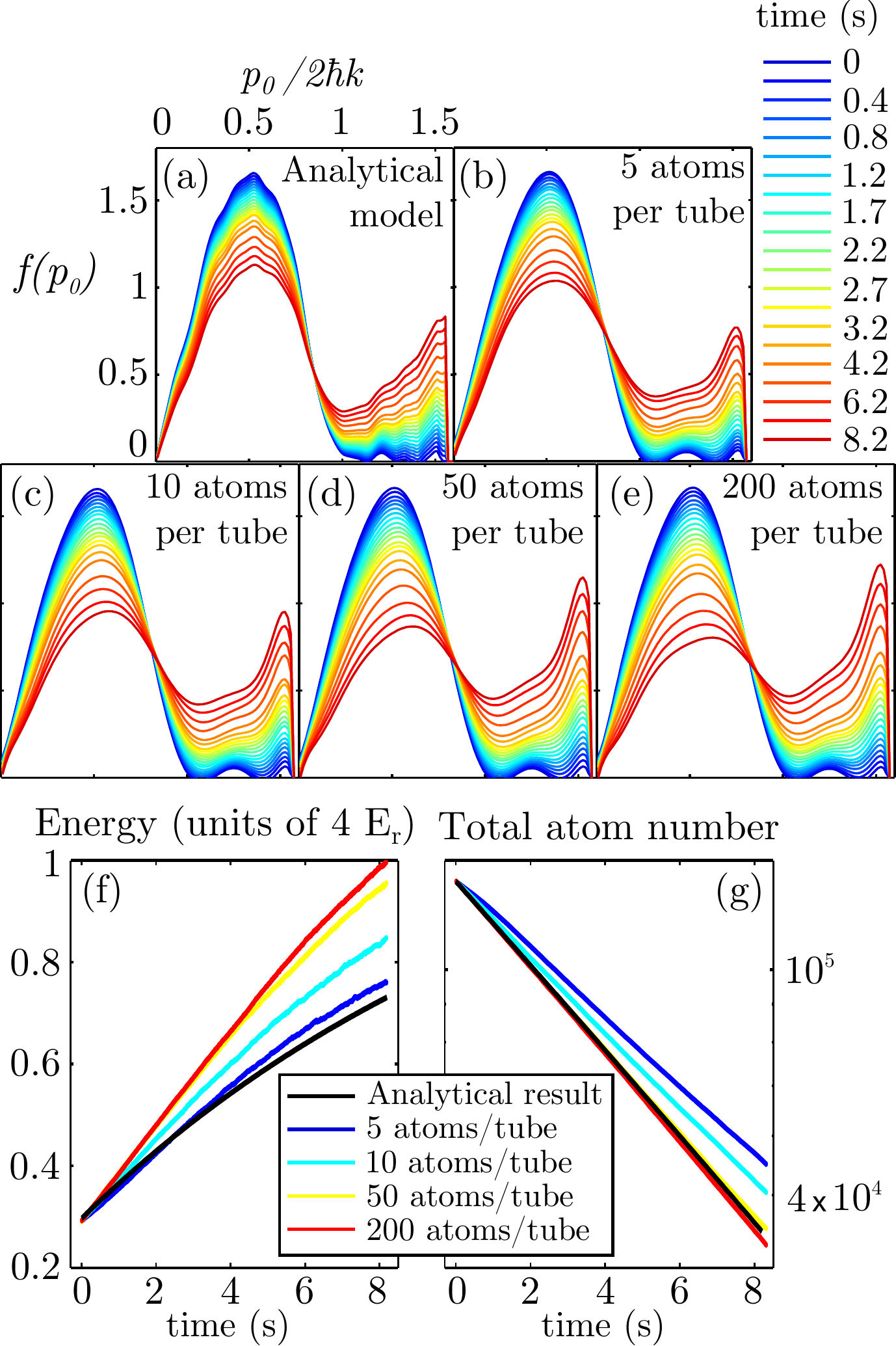}
\caption{Atom number dependence in the Monte-Carlo result. The result of the semi-analytical framework is compared to a Monte-Carlo approach in which only spontaneous emission processes and collisions are simulated. The evolution of the momentum distribution $f(p_0)$ is calculated by: the analytical framework (Fig (a)), Monte-Carlo simulations with a uniform distribution of atom number per tube, where Figs. (b), (c), (d), and (e) correspond to 5, 10, 50 and 200 atoms/tube. For the different calculations, we have also plotted in Fig. (f) the energy per atom in $\ket{0,0}$, and in Fig. (g) the evolution of the total atom number in the system.}
\label{FigA}
\end{center}
\end{figure}
	
	We begin with a collection of 1D Bose gases in a $12\, \mathrm{E_r}$ deep, 4.5 THz blue-detuned, 2D optical lattice, axially confined by a Gaussian trap with depth $U_0 = 10\,\mathrm{E_r}$ and a waist $w_0 = 41.6 \,\mathrm{\mu m}$. A quantum Newton's cradle oscillation is excited \cite{ExperimentalWork1}. The 2D optical lattice is shallow enough that two-body collisions can excite atoms to higher vibrational levels and tunneling occurs on short time scales.  The atoms therefore thermalize in 3D within 400 ms.  At this point we reliably know the transverse distribution among tubes. The lattice depth is then increased to $40\,\mathrm{E_r}$, making the system 1D from the perspective of two-body collisions.  We then remove the more energetic part of the distribution by decreasing the axial depth to $1\,\mathrm{E_r}$, after which we increase it back to $10\,\mathrm{E_r}$. At this point we have $N_0=1.3 \times 10^5$ uncorrelated atoms (a reduction of 75 $\%$) spread out over many more tubes than our typical quantum Newton's cradle, and three-body collisions are negligible.
In this configuration, atoms are distributed among tubes in $x$, $y$, according to a gaussian distribution of r.m.s size $6 \mathrm{\mu m}$, which corresponds to 87 atoms per tube initially in the center tube. We take a measured distribution as the starting point for the simulation. We then compare the simulated results to the experimental results over 8.2 seconds of evolution.
 
\subsection{Comparison with the results of the analytical framework}
\subsubsection{Differences in the evolution}

To get some insight into the validity of the approach of the semi-analytical framework, we have compared its results with the results of four Monte-Carlo simulations. To provide a fair comparison, we have intentionally disabled in the Monte-Carlo simulations all features that cannot be included in the semi-analytical framework. Therefore all simulations considered for this comparison describe an evolution due only to spontaneous emission processes and collisions. This is exactly the scope of what the semi-analytical model aims to capture. The four Monte-Carlo simulations that we use only differ from each other by the initial uniform atom number per tube (either 5, 10, 50, or 200 atoms/tube) which is chosen to be identical in all tubes, for the purpose of this comparison. All other parameters of the calculations are those specified in section \ref{sec:paramexp}. 

In Figure \ref{FigA}, we have plotted the results of this comparison. Figure \ref{FigA}(a) corresponds to the evolution of the $f(p_0)$ distribution as calculated from the semi-analytical model, whereas figures \ref{FigA} (b) to (e) correspond to the Monte-Carlo simulations with increasing initial population per tube. The agreement between the results of both methods is qualitatively good, and the analytical model captures the essence of the process that is mainly responsible for the evolution of the system, i.e. collisional de-excitations from transverse states with collisions (as discussed in section \ref{sec:deeccol}). 

In Fig. \ref{FigA}(f), we have plotted the mechanical energy of the atoms remaining in $\ket{0,0}$, and find that the analytical model does a better job in capturing the behavior when there are fewer atoms per tube. However, if we consider only the total losses of the system (Fig. \ref{FigA}(g)), the semi-analytical model agrees much better with a simulation with 200 atoms per tube, and the disagreement grows as the atom number per tube decreases.

The semi-analytical model is number-independent. However, as shown by the simulations, this atom number dependence is not negligible and can only be accounted for by a microscopic approach such as a Monte-Carlo simulation.

\subsubsection{Atom number dependence at the microscopic level}
\label{sec:N}

Figure \ref{FigE} shows the atomic trajectories that matter for the evolution of the system (i.e. the trajectories of atoms carrying some transverse excitations). They depend strongly on the atom number per tube. At low atom number per tube (Fig. \ref{FigE}(a)), atoms carrying some transverse excitations experience many oscillations before having a chance to transfer their transverse excitation to another atom via a collision. At large atom number per tube (Fig. \ref{FigE}(b)), they can transfer their transverse excitation very frequently, so that the ``motion'' of the transverse excitation is not oscillatory anymore.
Such behavior is presumably responsible for the differences in distribution shape and energy that we observe in Fig. \ref{FigA}.

To quantitatively find the root cause of the difference of behavior of the Monte-Carlo simulations as the atom number per tube increases, we have kept track of the populations in each transverse state, as well as of the atom losses as a function of the states they are in when they are lost. At large atom number per tube, the fractional population in higher transverse states is small, because the absolute population is large, and de-excitation collisions happen more frequently. With a small fraction of atoms in the first two excited states, the process by which atoms are lost because they have been promoted by spontaneous emission to a transverse state with 3 quanta of energy in one direction $x$ or $y$ is very rare. However for low atom number per tube, the population in the transverse states where $n_x$ or $n_y$ is  equal to 2 is not small enough to prevent some atoms from being promoted via spontaneous emission to a state with 3 quanta in one direction. These atoms  are thus lost from the system, taking with them the energy that would have led to extra heating and losses for atoms in $\ket{0,0}$, if they had had a chance to experience a collision back to the ground state. The difference in the incidence of this process accounts for 80 $\%$ of the difference in the loss along the axis from 200 atoms/tube and 5 atoms per tube. The fact that this process is absent in the semi-analytical model and minimal in large atom number per tube Monte-Carlo simulations explains why the losses agree in those two cases.

\subsection{Comparison with the evolution observed experimentally and impact of the different features/processes on the evolution}

\subsubsection{Validation of the model with the experimental observation}

We compare in Figures \ref{FigB} (a),(b) the evolution of the experimental distribution described in section \ref{sec:paramexp} to a Monte-Carlo simulation taking into account all features/processes which are known to be present experimentally. We also quantify the contribution of each feature/process in the total evolution by enabling/disabling those in subsequent Monte-Carlo simulations (Figs. \ref{FigB}(c)-(h)). For all these Monte-Carlo simulations, the distribution of atoms among tubes follows the gaussian shape observed experimentally in section \ref{sec:paramexp}.

Comparing Figs. \ref{FigB} (a),(b), we find very good agreement between the experimental observation and the Monte-Carlo simulation. Moreover, by looking at Fig. \ref{FigB} (i) and (j) we see that the simulation captures very well both energy increases and losses measured experimentally. It is worth noting that this agreement is obtained without any free parameters in the Monte-Carlo model related to spontaneous emission, collision rates or trajectories. The pointing noise is the only empirically adjusted value, and it is kept within the range of measurement uncertainty of its value. As we will see in the next subsection, it has only a very small effect on average energy or atom loss.

\begin{figure}[H]
\begin{center}
\includegraphics[width=3.4in, angle=0]{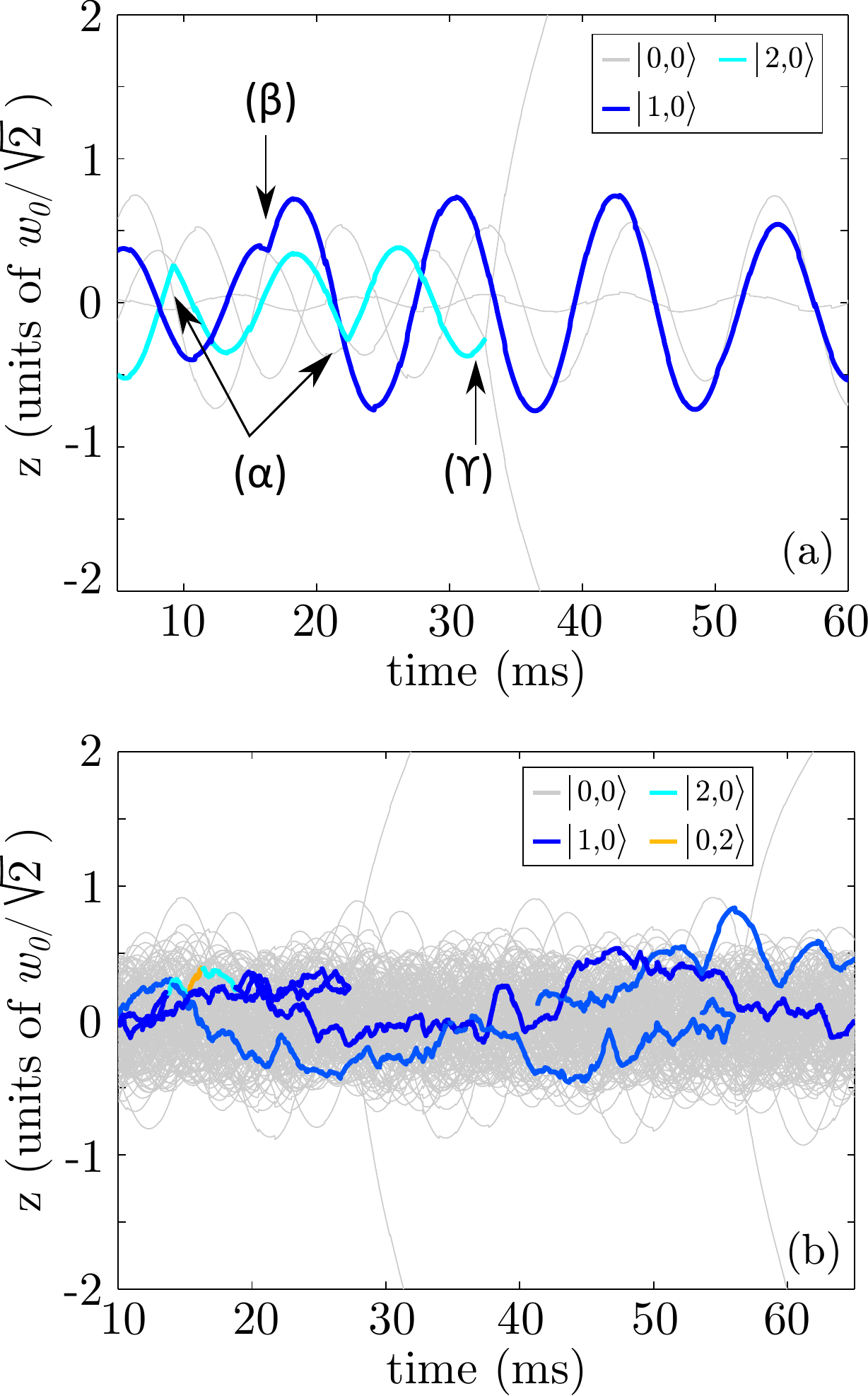}
\caption{Comparison of the atomic trajectories in a given tube with either 5 atoms (a) or 200 atoms (b). The transverse states of the atoms are color coded. In Fig. (a), we have indicated 3 collisional processes: $(\alpha) \,:\, \ket{2,0}+\ket{0,0} \rightarrow \ket{0,0}+\ket{2,0}$ (reflection), $(\beta) \,:\, \ket{1,0}+\ket{0,0} \rightarrow \ket{0,0}+\ket{1,0}$ (reflection),
$(\gamma) \,:\, \ket{2,0}+\ket{0,0} \rightarrow \ket{0,0}+\ket{0,0}$, (de-excitation leading to two losses in this example). In contrast to Fig. (a), the trajectories for the excited states in Fig. (b) do not exhibit complete oscillations in the axial trap, but a motion that looks more diffusive.  }
\label{FigE}
\end{center}
\end{figure}

\begin{widetext}

\begin{figure}
\begin{center}
\includegraphics[width=7in, angle=0]{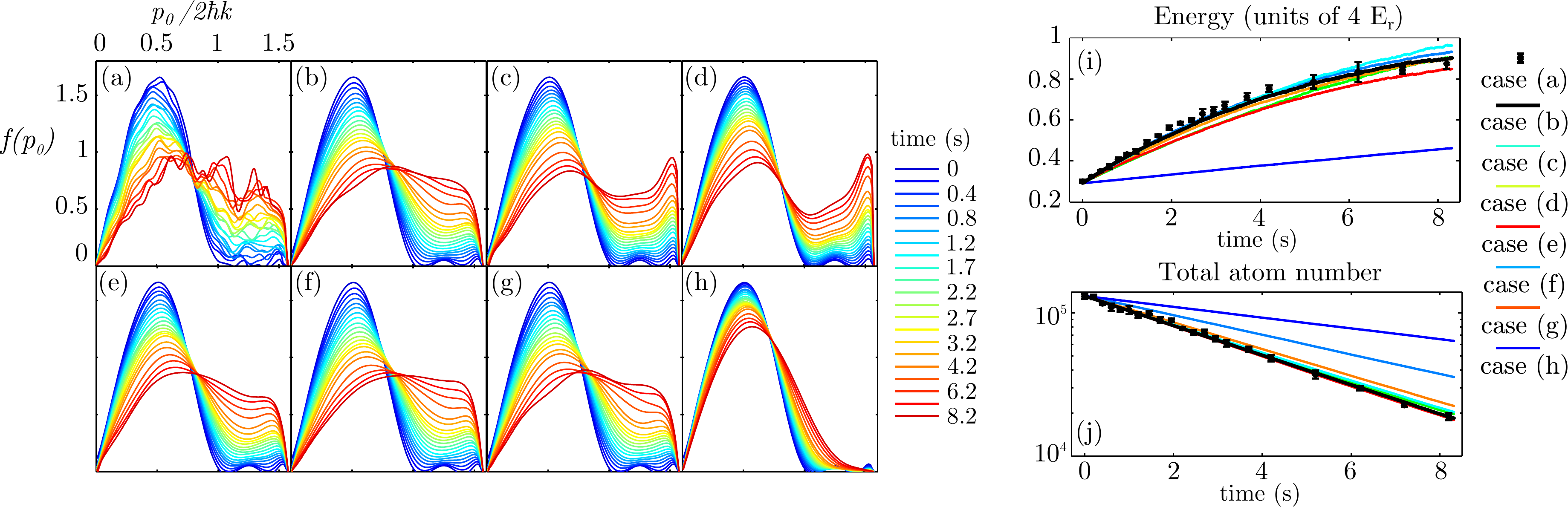}
\caption{Comparison (a) of the experimental observation of the distribution evolution, and (b) the Monte-Carlo simulation including all processes known to happen experimentally. We also show in figures (c) to (h) the contribution of different processes to the evolution, as given by Monte-Carlo simulations similar to (b) except that we have disabled the following features/processes:
(c) pointing noise ; (d) pointing noise and anharmonicity of lattice potential; (e) anharmonicity of lattice potential; (f) losses induced by background gas collisions; (g) spontaneous emission leading to a change of $F,m_F$ state; (h) all spontaneous emission processes. For the experimental evolution and the simulation results, we have also plotted a comparison of (i) the energy increase in $\ket{0,0}$, and (j) the atom population in the system as function of time.}
\label{FigB}
\end{center}
\end{figure}

\end{widetext}

We will now enumerate the features of the system that the Monte-Carlo simulation includes but that the semi-analytical model does not. By running the Monte-Carlo with or without these features/processes, we are able to quantify their impact on the total evolution.

\subsubsection{Experimental pointing noise of the axial potential}

The axial potential $U(z)$ is created experimentally with an optical beam which exhibits some pointing noise \cite{Gehm98}. This noise has been measured to be compatible with a value of $8.5 \times 10^{-6}\, \mathrm{\mu m^2/(rad \times s^{-1})}$ in the range of relevant axial trapping frequencies (less than or equal to 100 $\mathrm{Hz}$). In Fig.
\ref{FigB} (c), we have run the simulation without including this noise, whereas the simulation in Fig. \ref{FigB} (b) includes it. We observe that this noise has its main impact for large $p_0$, i.e. for atoms spending a lot of time near the edge of the the trap. Indeed, as can be seen in Fig. \ref{FigB} (i) and (j), pointing noise is responsible for a reduction of the net heating in $\ket{0,0}$ and for some slight increase of the losses. Despite the fact that the axial trap is anharmonic here, these observations are still compatible with the simple picture of an axial harmonic trap where pointing noise heats up all atoms equivalently independently of their energy, but due to the finite depth of the trap leads to the loss of atoms much more energetic than the average, thus resulting in net cooling.

\subsubsection{Anharmonicity of the lattice potential}

\label{sec:anharm}

In the analytical model, we treat the processes $\ket{1,0}+\ket{1,0}$ and $\ket{2,0}+\ket{0,0} \rightarrow \ket{0,0}+\ket{0,0}$ equivalently. 

In the harmonic approximation, for a given relative momentum, the probability to decay is the same for $\ket{1,0}+\ket{1,0}$ and $\ket{2,0}+\ket{0,0} \rightarrow \ket{0,0}+\ket{0,0}$ (see Fig. \ref{PABGD}). Although this does not strictly hold for an anharmonic potential, the experiment is insensitive to the exact values of the collision rates. For instance, if the de-exciting collisions were a little less frequent, then a slightly larger fraction of atoms would be shelved in the transverse excited states. But that is a transient effect that barely changes the steady state heating rate. We therefore do not attempt anharmonic corrections to the rates in the Monte-Carlo simulations.

However, the energy $E_g$ transferred to the axial motion by these processes is identical only in the limit where the transverse potential is purely harmonic, which is not the case when the transverse sinusoidal potential is created by an optical lattice. In this case, the energy spacing $\Delta E_{n+1,n}$ between consecutive transverse levels decreases as the transverse quantum number $n$ increases. Even in the limit of an infinite lattice depth, for all $n \geq 0$, $\Delta E_{n+1,n}=\Delta E_{n+2,n+1}+E_r$. Thus for any lattice depth the energy released by a collision $\ket{1,0}+\ket{1,0}\rightarrow \ket{0,0}+\ket{0,0}$ is higher than the energy released by $\ket{2,0}+\ket{0,0}\rightarrow \ket{0,0}+\ket{0,0}$ by at least 1 $E_r$. However, if the lattice depth is large enough, the {\it relative} difference in energy can be negligible. For example for a lattice depth of 40 $E_r$ in the quantum Newton's cradle experiment \cite{ExperimentalWork1}, the energy released by the first collision process is 23.1 $E_r$ and the energy released by the second one is 21.8 $E_r$. By assuming that both processes release the average energy 22.5 $E_r$, we make a relative error of less than $3\%$ for the energy transferred to the axial motion, which makes little difference.

In contrast, the process $\ket{1,0}+\ket{1,0} \rightleftharpoons \ket{2,0}+\ket{0,0}$, which does not modify the momentum distribution in the harmonic case, leads to an energy exchange of order $1 E_r$ in the anharmonic case. By inducing even this small energy exchange, this process can significantly affect the evolution of the momentum distribution at any lattice depth. The amount of energy deposited is comparable to the energy deposited during a de-excitation collision. But it has a much larger effect per energy deposited on momentum distributions. As illustrated in Fig. 2, de-excitations to the ground state redistribute affected atoms across the whole $p_0$ distribution. In contrast, the collisions described here act more like a smoothing function that slightly spreads out an affected atom's $p_0$, tending to erase narrow $p_0$ features.

In the Monte-Carlo simulation of Fig. \ref{FigB} (d), we have intentionally set the values $\Delta E_{n+1,n}$ to be constant and equal to $\Delta E_{1,0}$, so that it simulates an harmonic transverse potential with transverse levels equally spaced. This simulation is also performed without any pointing noise for clarity. Comparing Fig. \ref{FigB} (d) to (c), we can see the modifications induced by the inclusion of anharmonicity alone. As discussed above, the small energy exchange of the process $\ket{1,0}+\ket{1,0} \rightleftharpoons \ket{2,0}+\ket{0,0}$ is responsible for a broadening of the distribution particularly visible for $p_0/2\hbar k$ between 1 and 1.5. It is also responsible for some part of the net heating as seen on Fig. \ref{FigB} (i). However, as expected, its impact on the losses is minimal (Fig. \ref{FigB} (j)). 

If we look at the impact of anharmonicity on a simulation including pointing noise, i.e. if we compare Fig. \ref{FigB} (e) to (b) the effect is less visible, but still there. It is merely hidden by the bigger effect of the pointing noise.

\subsubsection{Impact of background gas collisions}

Experimentally, there is a finite lifetime due to losses induced by collisions with the background gas.
The loss rate measured with an atomic cloud in a deep 3D trap is 0.087 Hz. 
In Fig. \ref{FigB} (f), we show the simulation where we have ignored loss; it is very close to the full simulation (Fig. \ref{FigB} (b)). In Fig. \ref{FigB} (i) and (j), we observe  larger losses and slightly less net heating when this source of losses is included. The change in heating is merely explained by the previous observation (section \ref{sec:N}) that tubes with small atom number see a smaller net heating than tubes with a larger atom number.
  
\subsubsection{Spontaneous emission induced hyperfine state change}

As described in \cite{TransitionRates}, for a non-infinite optical lattice detuning there is some probability that a spontaneously emitted photon from an atom leads to a change of hyperfine state. This can lead to complicated dynamics when the atom is no longer trapped or is differently trapped, as in Ref. \cite{ExperimentalWork1}, where only atoms in $\ket{F=1,m_F=1}$ are magnetically levitated.  When unlevitated atoms accelerate on their way out of the cloud, they collide with trapped atoms, possibly causing energy deposition and possibly extra loss. 

At our typical experimental detuning (4.5 THz), we can see that the effect of these processes is minimal by comparing Fig. \ref{FigB} (g) and (b). The difference in loss rates between both simulations is 0.0235 Hz, which is also the total transition rate to untrapped states as calculated {\it ab initio} in \cite{TransitionRates}. This shows that at this density and detuning, an unlevitated atom does not induce significant extra losses by colliding on its way out of the trap.

\section{Conclusion}

We have presented a general formalism for calculating energy deposition due to the spontaneous emission of optical lattice light. For arbitrary dimension lattices of any detuning and polarization, as long as they are sufficiently deep and their detuning exceeds hyperfine structures, the formalism allows one to keep track of how the atoms' internal and external states change, and how these various changes are correlated with one another. As an example, we have calculated energy deposition for $^{87}$Rb atoms in linearly polarized 2D optical lattices. That particular case is important for understanding how quantum Newton's cradle momentum distributions evolve due to spontaneous emission.

\section*{Acknowledgments}
This work was supported by the NSF (PHY 11-02737), the ARO, and DARPA.

\appendix
\addappheadtotoc

\section{Select derivations}
\label{AppGpp0}
\subsection{Derivation of Eq. \ref{eq:Gpp0}}

\subsubsection{General expression}
Let us consider a particle with a mechanical energy $p_0^2/2m$ oscillating between the points $z_1$ and $z_2$ in the potential $U(z)$. We have, at any point $z$
\begin{equation}
\frac{p_0^2}{2m}=\frac{p^2}{2m}+U(z)
\label{eqmec}
\end{equation}
and the oscillation period $T(p_0)$ of this particle is then given by
\begin{equation}
T(p_0)=2\int_0^{T/2} dt=2m\int_{z_1}^{z_2}\frac{dz}{\sqrt{p_0^2-2mU(z)}}
\end{equation}
Using the equation of motion
\begin{equation}
\frac{dp}{dt}=-\frac{dU}{dz}
\end{equation}
we obtain that, averaged on half a cycle ($0\leq p \leq p_0$), the probability $G(p,p_0)dp$ to find the atom at a momentum between $p$ and $p+dp$ is
\begin{equation}
G(p,p_0)dp=\frac{dt}{T/2}=2\frac{dp}{T}\sum_{z_i}\frac{1}{\left|\frac{dU}{dz}\right|_{z_i(p,p_0)}}
\end{equation}
where the sum is taken on all the positions $z_i$ satisfying eq. \ref{eqmec} for the considered $p$.
In the following, we express $G(p,p_0)$ as a function of $p,p_0$ in two specific cases.

\subsubsection{Case of an harmonic trap}

In the case of an harmonic trap
\begin{equation}
\frac{dU}{dz}=m\omega^2z=\omega\sqrt{p_0^2-p^2}
\end{equation}
and
\begin{equation}
G(p,p_0)=\frac{2}{\pi\sqrt{p_0^2-p^2}}
\label{eq:Gpp0harm}
\end{equation}

\subsubsection{Case of a gaussian trap}

In the case of a gaussian trap of depth $U_0$ and waist $w_0$,
\begin{equation}
U(z)=U_0\left[1-\exp\left(-2 z^2/w_0^2\right)\right].
\label{eqUzapp}
\end{equation}
By inverting Eq. \ref{eqUzapp} we find
\begin{equation}
z=\pm w_0\sqrt{\frac{\ln \left[U_0/(U_0-U(z))\right]}{2}}.
\end{equation}
We can write
\begin{eqnarray}
\frac{dU}{dz}=&\frac{4U_0z}{w_0^2}\exp\left(-2 z^2/w_0^2\right)\\
=& \frac{4z}{w_0^2}\left[U_0-U(z)\right]\\
=& \pm \frac{2\sqrt{2}}{w_0} \sqrt{\ln \left[U_0/(U_0-U(z))\right]}  \left[U_0-U(z)\right].
\end{eqnarray}
By using $U(z)=p_0^2/2m-p^2/2m$ and by introducing the reduced units $\tilde{p}$ and $\tilde{p}_0$ defined in section \ref{refp0tilte}, we obtain Eq. \ref{eq:Gpp0}.

\section{Collision rates and probability function}

The aim of this appendix is to derive the expression of the probability $g_{\mathrm{rel}}(p,p_0,k,t) dk$ that an atom with a given $p_0$ and $p$ experiences a collision $\alpha+\beta \rightarrow \gamma+\delta$ involving a relative momentum between $k$ and $k+dk$. This function appears naturally when we express the collision rates in $p_0$ representation. All the functions and variables can have an explicit time dependence. We drop this dependence in the following to lighten the notations.

In order to derive the rates of vibrational state changing collisions, we consider a single tube containing $N$ atoms in different transverse levels $\alpha$,
$N=\sum_{\alpha} N_{\alpha}$. Collision rates are proportional to $n_{\alpha}(p,z)$, the probability density of an atom in the transverse excited state $\alpha$ with momentum $p$ at position $z$, which is related to $N_{\alpha}$ according to $\int\!dp\,dz\, n_{\alpha}(p,z)= N_{\alpha}$. Collision rates also depend on the probability density of the collision partner, $n_{\beta}(p,z)$ and the colliding atoms' relative momentum $2k$ in the lab frame. Specifically, the collision rate in $(p,z)$ representation is given by
\begin{equation}
\Gamma_{\alpha \beta}^{\gamma \delta}(p,z)=\\\!\!\!\left[\int\!\!dk \frac{2|k|}{m} n_{\beta}(p',z\!) P_{\alpha \beta}^{\gamma \delta}(k)\right]n_{\alpha}(p,z).
\label{colratepz}
\end{equation}
where $p'=p+2k$ is the momentum of atom $\beta$.

Since $p_0$, not $p$ or $z$, is the conserved quantity in the free evolution of the 1D gas, it is useful to transform the rates in Eq. \ref{colratepz} into $p,p_0$ space. Observing that each set $(p,p_0)$ corresponds to two sets $(p,z)$, we perform the change of variable, $z\!\rightarrow \!p_0$, by introducing ${p'}_0^2/2m$, the mechanical energy of atom $\beta$ before the collision, and using conservation of energy, ${p'}_0^2-{p'}^2=p_0^2-p^2 =2mU(z)$. We finally get the $(p,p_0)$ representation of the rate of eq. \ref{colratepz}
\begin{equation}
{}^{}\!\!\!\Gamma_{\alpha \beta}^{\gamma \delta}(p,p_0)\!\!=\!\!\Biggl|\frac{dU}{dz}\Biggr|_{p_0,p}\!\!\!\!\!\!\!\!\!\int\!\!\!dk\frac{|k|}{p_0'} n_{\beta}(p',p_0') P_{\alpha \beta}^{\gamma \delta}(k)n_{\alpha}(p,p_0).
\label{colratepp0}
\end{equation}
The $(p,p_0)$ space is a valid description of the collisions as long as we assume that an atom with a given mechanical energy (i.e. a given $p_0$) samples the whole $z$ space. This is not always true, and depends on the density, as seen in the Monte-Carlo simulations (see Fig. \ref{FigE}). However, without keeping track of every single trajectory, it is the best assumption we can make.

Monte-Carlo simulations show that the two body collisions between different transverse states rapidly ensure that all levels have the same momentum distribution. By introducing the common normalized momentum distribution $d$ defined for all states $\alpha$ as $d(p,p_0)=n_{\alpha}(p,p_0)/N_{\alpha}=G(p,p_0)f(p_0)$, we can write the rate in Eq. \ref{colratepp0} as the product of a collision constant $\kappa_{\alpha \beta}^{\gamma \delta}$ and the populations of the incoming colliding atomic states,
\begin{equation}
\Gamma_{\alpha \beta}^{\gamma \delta}(p,p_0)= \kappa_{\alpha \beta}^{\gamma \delta}(p,p_0) N_{\beta}N_{\alpha},
\label{eqGammaNiNj}
\end{equation}
where
\begin{eqnarray}
\kappa_{\alpha \beta}^{\gamma \delta}(p,p_0)&=&\left[\int \!\!dk f_{\mathrm{rel}}(p,p_0,k)\right]  d(p,p_0),
\label{eqkappaNiNj}\\
f_{\mathrm{rel}}(p_0,p,k)&=&\frac{d(p',p_0') P_{\alpha \beta}^{\gamma \delta}(k) |k|}{p_0'} \left|\frac{dU}{dz}\right|_{p0,p}.
\label{frel}
\end{eqnarray}

The probability function $g_{\mathrm{rel}}(p_0,p,k)$, that an atom with a given $p_0$ and $p$ experiences a collision involving a relative momentum $k$ is given by $f_{\mathrm{rel}}$ normalized,
\begin{equation}
{}^{}\!\!\!g_{\mathrm{rel}}\!=\!\frac{N_{\beta}f_{\mathrm{rel}}}{\int\!dk N_{\beta} f_{\mathrm{rel}} }
\!=\!\frac{d(p'\!,\!p_0') P_{\alpha \beta}^{\gamma \delta}(k) |k|/p_0'}{\int\!dk \, d(p'\!,\!p_0') P_{\alpha \beta}^{\gamma \delta}(k) |k|/p_0'}
\label{grel}
\end{equation}

We can apply the previous formulas to calculate the collision rate averaged over a simple momentum distribution in the analytical case of the harmonic oscillator.
We consider two atoms, one in state $\alpha$, the other one in $\beta$ such that $N_\alpha=N_\beta=1$. We consider that each atom has maximum momentum $p_a$, such that
\begin{equation}
n_{\alpha,\beta}(p,z)=\frac{\delta\left(z-\frac{\sqrt{p_a^2-p^2}}{m\omega}\right)+\delta\left(z+\frac{\sqrt{p_a^2-p^2}}{m\omega}\right)}{\pi\sqrt{p_a^2-p^2}}
\end{equation}
We obtain the collision rate averaged over $p$ and $z$
\begin{equation}
\Gamma_{\mathrm{avg}}=\int dp\, dz\, \Gamma_{\alpha \beta}^{\gamma \delta}(p,z)= \frac{\omega}{\pi} \int_{0}^{p_a} dp \, P_{\alpha \beta}^{\gamma \delta}(p)\, G(p,p_a)
\end{equation}
with $G(p,p_a)$ given by Eq. \ref{eq:Gpp0harm}.
Assuming no dependence on $k$ of the collision, we obtain $\Gamma_{\mathrm{avg}}=\omega/\pi$, i.e. that there are two collisions per oscillation period, as expected. Of course, we obtain the same result if we use the $(p,p_0)$ representation with $n_{\alpha,\beta}(p,p_0)=\delta(p_0-p_a)\,G(p,p_0)$.


\begin{thebibliography}{99}
\bibitem{footnoteN1} The ratio $N_1/N_0$ is much less in the case of quantum Newton's cradle distributions \cite{ExperimentalWork1} with a large atom number.
\end{thebibliography}
\end{document}